\documentclass{jaa}
\usepackage{natbib}
\usepackage{comment}
\usepackage{graphicx}
\usepackage{verbatim}
\usepackage[colorlinks, allcolors=blue]{hyperref}
\usepackage{adjustbox}
\usepackage{multirow}
\usepackage{multicol}
\usepackage[flushleft]{threeparttable}
\usepackage[utf8]{inputenc}
\usepackage{subfigure}
\usepackage{subcaption}
\usepackage[l3]{csvsimple}
\usepackage{booktabs}         
\usepackage{amsmath}           
\usepackage{siunitx}     
\usepackage{float}
\usepackage{longtable}
\usepackage{stfloats}
\usepackage{appendix}

\begin{document}\sloppy

\title{Exploration of UV-bright sources in Globular cluster NGC 3201 using UVIT observations}


\author{Manasi Kulkarni\textsuperscript{1}, Sonika Piridi\textsuperscript{1}, Ranjan Kumar\textsuperscript{2}, Ananta C. Pradhan\textsuperscript{1*}}

\affilOne{\textsuperscript{1} Department of Physics and Astronomy, National Institute of Technology, Rourkela, Odisha - 769008, India.\\}
\affilTwo{\textsuperscript{2} Department of Physics, U. R. College, Rosera, A constituent unit of Lalit Narayan Mithila University, Darbhanga, Bihar - 848210, India\\}

\twocolumn[{
\maketitle
\corres{acp.phy@gmail.com}
\vspace{0.5cm}

\begin{abstract}
We present a comprehensive study of the ultraviolet (UV)-bright stellar population in the globular cluster NGC 3201 using the Ultraviolet Imaging Telescope (UVIT) onboard AstroSat. By combining far-UV photometry of the UVIT F148W and F169M filters with Gaia and ground-based telescope optical data, we construct UV$-$optical color-magnitude diagrams (CMDs) to identify UV-bright stellar evolutionary phases. Physical parameters of all detected sources are derived by fitting stellar atmosphere models to the observed multi-wavelength spectral energy distributions (SEDs). The effective temperatures of blue horizontal branch stars (BHBs) span the range 7250$-$11250 K.  We identify two extreme horizontal-branch stars (EHBs) with effective temperatures of approximately 25,000 K and 17,000 K; one lies within the cluster core radius ($r_\mathrm{c}$=1.3$'$), while the other is located outside. We also detect two candidate ``gap'' objects in the transitional phase between HB stars and white dwarfs, which may represent hot blue hook (BHk) stars. For the blue straggler stars (BSSs), we determine effective temperatures, radii, and luminosities through SED fittings, and estimate their masses and isochrone-equivalent ages using PARSEC isochrones. Finally, we studied the normalized cumulative radial distribution of the observed BHB stars, BSSs, MS stars, RGB stars, and variable stars along with the specific frequency of BSSs across the cluster. These results provide new insights into the late-stage stellar evolution and dynamical processes in NGC 3201.
\end{abstract}

}]


\doinum{}
\artcitid{\#\#\#\#}
\volnum{000}
\year{0000}
\pgrange{1--}
\setcounter{page}{1}
\lp{1}

\section{Introduction}\label{sec:intro}
Globular clusters (GCs) are among the oldest stellar systems in the Universe, serving as fossil records of early star formation, chemical evolution, and dynamical processes in galaxies \citep{2006ARA&A..44..193B}. These dense, spherical collections of hundreds of thousands to millions of stars are typically characterized by ages exceeding 10–13 Gyr, low metallicities, and complex internal structures shaped by multiple stellar populations \citep{Bastian2018}. The diameter of GCs is a few arcminutes, and the average separation between the stars within a GC is approximately one light year. In recent decades, ultraviolet (UV) observations have revolutionized our understanding of GCs by revealing hot, evolved stellar populations that dominate the UV flux but remain inconspicuous at optical wavelengths \citep{Dalessandro2012, Nardiello2018, Piridi2026}. These UV-bright sources provide critical insights into stellar evolution pathways, binary interactions, mass transfer, and the formation mechanisms of exotic objects such as blue straggler stars (BSSs), extreme horizontal branch (EHB) stars, and white dwarfs (WDs) \citep{Sahu2019ApJ, Sahu2019MNRAS, Prabhu2021ApJ, Kumar2021MNRAS, Rani2021ApJ, Kumar2022MNRAS, Vaidya2022MNRAS, Jadhav2023ApJ}.

The UV region is particularly sensitive to stars with effective temperatures above $\sim$ 7000$-$8000 K. In old populations like GCs, the far-UV (FUV: 1300$–$1800 \AA) and near-UV (NUV: 2000$–$3000 \AA) light primarily arises from blue horizontal branch (BHB) stars, post-asymptotic giant branch (post-AGB) stars, BSSs, hot subdwarfs, cataclysmic variables (CVs), and low-mass WDs \citep{Rey2007ApJS, Dalessandro2012AJ, Schiavon2012AJ, Sahu2019MNRAS, Prabhu2021ApJ, Kumar2022MNRAS, Piridi2026}. In the UV wavelength regime, these hot stars dominate over the cooler stars, making them much easier to identify even in the densely crowded central regions of GCs. Consequently, UV imaging is an exceptionally powerful tool for studying the formation and evolution of hot evolved stars in GCs \citep{Moehler2019}. These objects trace advanced evolutionary stages and dynamical interactions. For instance, BHB stars are core helium-burning stars that have lost significant envelope mass and appear blueward of the main sequence (MS) turn-off in color-magnitude diagrams (CMDs). Their distribution and temperature spread help resolve the long-standing ``second parameter'' problem in HB morphology, where clusters of similar metallicity exhibit vastly different HB extensions \citep{Rey2007ApJS}. BSS, brighter and bluer than the main-sequence turn-off, are thought to form via stellar collisions or binary mass transfer, offering a window into the dynamical history of the cluster \citep{FusiPecci1992AJ,Bailyn1995ARA&A,Piotto2004ApJ,Mapelli2006MNRAS,Ferraro2009Nature,Gratton2010AARv}.

Recent UV investigations of GCs using space-based observatories such as \textit{Galaxy Evolution Explorer (GALEX), Hubble Space Telescope (HST),} and Ultraviolet Imaging Telescope (UVIT) have significantly advanced our understanding of hot and evolved stellar populations in GCs. {\it GALEX} studies revealed that the integrated UV properties of GCs are strongly influenced by HB morphology, helium enrichment, and the presence of EHB stars, helping to address the long-standing second-parameter problem in cluster evolution \citep{Rey2007ApJS, Schiavon2012AJ,Dalessandro2012AJ}. High-resolution UV imaging from {\it HST} further enabled the identification of BSSs, CVs, WDs, and post-AGB stars in crowded cluster cores, thereby constraining stellar collision rates, binary evolution, and dynamical interactions in GCs \citep{Knigge2002ApJ, Piotto2004ApJ, Ferraro2009Nature}. More recently, UVIT onboard AstroSat has provided deep FUV and NUV observations with improved spatial resolution, enabling detailed characterization of UV-bright stellar populations in GCs such as NGC 2808, NGC 288, NGC 1261, NGC 4590, and NGC 4147 \citep{Sahu2019MNRAS, Prabhu2021ApJ, Rani2021MNRAS, Kumar2021MNRAS, Kumar2022MNRAS}. Integrated UV analyses using UVIT have additionally demonstrated correlations between FUV emission, metallicity, and HB extension, reinforcing the importance of UV observations in probing the evolution history and stellar content of old stellar systems \citep{Bianchi2011ApSS, Pandey2021MNRAS, Piridi2026}.

NGC 3201 (Coordinates: $\alpha_{2000}$ = 10${\mathrm{h}}$ 17${\mathrm{m}}$ 37${\mathrm{s}}$, $\delta_{2000}$ = $-46^{\circ}$ 24$^{\prime}$ 40$^{\prime\prime}$), also known as Caldwell 79, is a Galactic GC located in the constellation Vela at a heliocentric distance of 4.9 kpc \citep{Harris1996}. It is a metal poor cluster with metallicity of $[\mathrm{Fe}/\mathrm{H}] = -1.46 \pm 0.15 \, \mathrm{dex}$ \citep{Kravtsov2009} and alpha enhancement of $[\alpha/\mathrm{Fe}] = 0.37 \pm 0.04$ \citep{Magurno2018}. It is a massive cluster with an age $\sim 11.85 \pm 0.74$ Gyr \citep{Ying2024}. Recent studies using Gaia, {\it HST}, and high-resolution spectroscopy have revealed several remarkable features about the cluster. The cluster possesses a high radial velocity ($\sim$490 km/s) and a retrograde orbit, strongly suggesting an accreted origin from a disrupted dwarf galaxy. It hosts multiple stellar populations, with the second-generation (enriched) stars being more centrally concentrated and showing distinct kinematic properties compared to the primordial population \citep{Munoz2013}. 

Notably, NGC 3201 has gained significant attention for its rich binary population and the presence of stellar-mass black holes. The first dynamical detection of a stellar-mass black hole in a GC was reported in NGC 3201 using Multi Unit Spectroscopic Explorer (MUSE) spectroscopy \citep{Giesers2018}. BSSs in the cluster show chemical abundances consistent with the host cluster and exhibit high binary fractions, supporting formation through mass transfer and stellar collisions \citep{Giesers2019}. Gaia data have also revealed an extensive tidal stream stretching over $\sim 140^{\circ}$ on the sky, confirming ongoing tidal disruption by the Milky Way \citep{Palau2021}.

In this paper, we have studied the hot evolved sources of the cluster using UVIT observations. The paper is organized as follows. In Section ~\ref{sec:Observation and Data Reduction}, we present the observation and data reduction, and in Section ~\ref{sec: Cluster parameter}, we estimate the cluster distance. In Section ~\ref{sec:CMD}, we present UV and optical CMDs. Section ~\ref{sec:sed fittings} investigates different stellar populations through SED fitting, and Section ~\ref{sec:Radial distribution} is focused on the radial distribution of BHB, BSS, MS, RGB, and variable stars. All magnitudes given in this work are in the AB system. 

\section{Observation and Data Reduction} \label{sec:Observation and Data Reduction}

The cluster NGC 3201 was observed in two UVIT FUV filters: F148W ($\mathrm{\lambda_{eff}}$ = 1481 \AA) and F169M ($\lambda_\mathrm{eff} =$ 1608 \AA). The raw level-1 data were processed using the CCDLAB pipeline \citep{ccdlab2017, ccdlab2021}. After applying all the necessary corrections, the individual frames were registered and merged to produce a final image with an effective exposure time of 6705 (6224) sec for F148W (F169M) filter. We use the automated astrometry technique in the CCDLAB to perform the astrometry of the image using the Gaia DR3 data \citep{gaiadr3}. The WCS solution uncertainty is 0.1$''$ in both filters. 

After obtaining the science-ready images, we estimated the sky background by fitting boxes of size 21 $\times$ 21 pixels, as described in \cite{2024uvitdr1}. The estimated mean sky background is 0.5 (0.2) counts/pixel for F148W (F169M) filters. We use source-extractor \citep{1996Bertin} to extract the sources using the estimated background values and IRAF to perform the PSF photometry of the sources. The PSF FWHM of the image was computed to be 3.8 pixels ($\sim$ 1.6$''$) and 3.6 pixels ($\sim$ 1.5$''$) for F148W and F169M filters, respectively. PSF magnitudes are corrected for apertures and saturation effects. Finally, we obtained 200 and 203 sources in F148W and F169M filters, respectively. The maximum uncertainty in the magnitudes is 0.2 mag. A sample of the photometry file is given in \autoref{app:online_tables}. The complete table will be available online.
The magnitudes were corrected for extinction using the extinction law of \citet{Cardelli1989} and assuming a constant extinction throughout the cluster. The E(B$-$V) value for the cluster is 0.219 \citep{Schlafly&Finkbeiner2011}. 

\begin{table*}
\centering
\caption{
Photometry source catalog of NGC 3201 observed in F148W and F169M filters. The UVIT counterpart in the HST and Gaia catalog are indicated in the ''match" column. The complete catalog is available online at \url{https://doi.org/10.5281/zenodo.21758849}.}
\label{app:online_tables}
\begin{threeparttable}    
    \begin{tabular}{|r|r|r|r|r|r|l|c|c|}
    \hline
      \multicolumn{1}{|c|}{RAJ2000} &
      \multicolumn{1}{c|}{DECJ2000} &
      \multicolumn{1}{c|}{F148W} &
      \multicolumn{1}{c|}{e\_F148W} &
      \multicolumn{1}{c|}{F169M} &
      \multicolumn{1}{c|}{e\_F169M} &
      \multicolumn{1}{c|}{match} &
      \multicolumn{1}{c|}{Member} &
      \multicolumn{1}{c|}{Type} \\
    \hline
        154.15599 & -46.55559 & 22.39 & 0.16 & 22.27 & 0.2 & Gaia & 1 & 1\\
      154.20232 & -46.40032 & 19.51 & 0.03 & 19.16 & 0.04 & Gaia & 1 & 1\\
      154.47343 & -46.4403 & 18.56 & 0.02 & 18.29 & 0.03 & Gaia & 1 & 1\\
      154.43745 & -46.42044 & 21.17 & 0.06 & 20.93 & 0.1 & HST & 1 & 2\\
      154.41533 & -46.42253 & 22.52 & 0.15 & 999.0 & 999.0 & HST & 1 & 2\\
    
      \hline
    \end{tabular}
   \begin{tablenotes}
    \item {\bf match:} HST, Gaia, Unknown; {\bf Member:} 0=  non cluster member, 1= cluster member, and -1 = unknown; {\bf Type:} 0= post-HB, 1 = HB, 2 = BSS, and -1 = unknown.
   \end{tablenotes}  
 \end{threeparttable} 

\end{table*}

\section{Distance Estimation of the Cluster} \label{sec: Cluster parameter}
\begin{figure}
    \centering
    \includegraphics[width=\columnwidth]{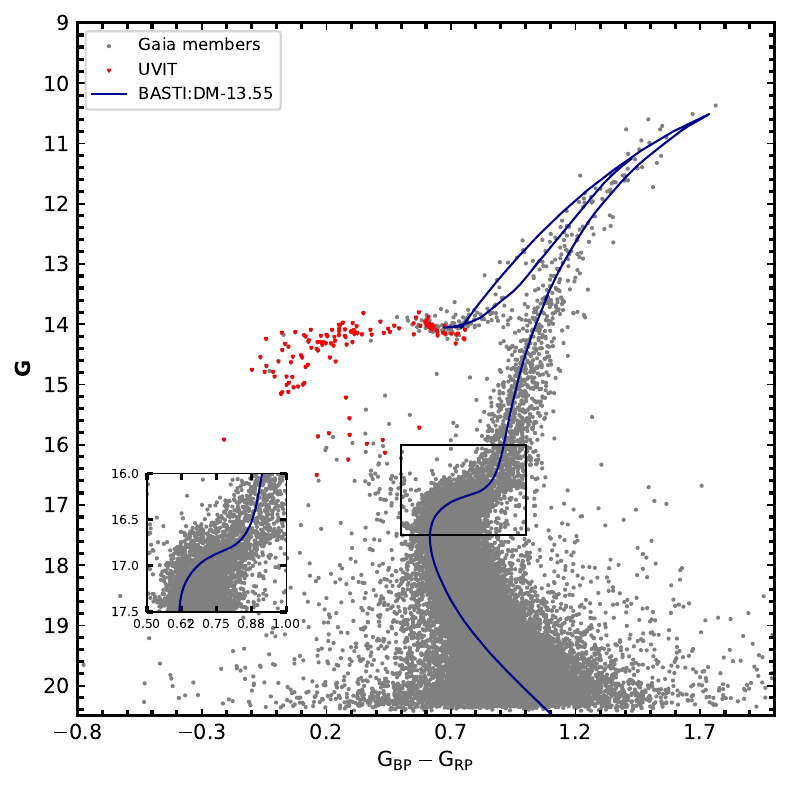}
    \caption{CMD of the GC in the Gaia photometric systems, showing $G_{\mathrm{BP}}$ - $G_{\mathrm{RP}}$ versus G. The UVIT detections are plotted as red triangles. Overplotted are BaSTI-IAC isochrones corresponding to age $=$ 11.80 Gyr, [Fe/H] $= -$1.40 dex and DM = 13.55.}
    \label{fig:distance_modulas}
\end{figure}

The Fig.\ref{fig:distance_modulas} shows the $\rm G_{BP} - G_{RP}$ versus G color-magnitude diagram (CMD) diagram of NGC 3201, constructed using 32464 stars selected within the tidal radius of the cluster and satisfying $P_{\mathrm{member}} \geq 0.6$. We adopted isochrones of BASTI-IAC (Bag of Stellar Tracks and Isochrones)\footnote{\url{http://basti-iac.oa-abruzzo.inaf.it/index.html}} models \citep{Pietrinferni2021} for the cluster parameters of age = 11.80 Gyr and metallicity [Fe/H] = $-1.40$ dex. Isochrone fits were examined for multiple values of distance moduli (DM) ranging from 13.3 to 13.8 in steps of 0.1. 
The isochrone corresponding to DM of $13.55 \pm 0.05$ (5.13 $\pm$ 0.12 kpc) fits well across the main sequence turn off (MSTO) as well as the SGB and RGB stars. Our derived distance of 5.13 $\pm$ 0.12 kpc is in good agreement with previous studies and falls well within the range reported in the literature \citep{ArellanoFerro2014, Monty2018, Baumgardt2021}.
 
\section{UV--Optical CMDs and identification of UV-bright stars} \label{sec:CMD}
After compiling the list of UVIT sources in the globular cluster NGC 3201, we cross-matched our observed catalog with the HST catalog in the inner region and the Gaia catalog in the outer region to identify cluster members. The identification and classification of different types of cluster members are described in the following subsections.

\begin{figure*}
  \centering
  \includegraphics[width=0.49\textwidth]{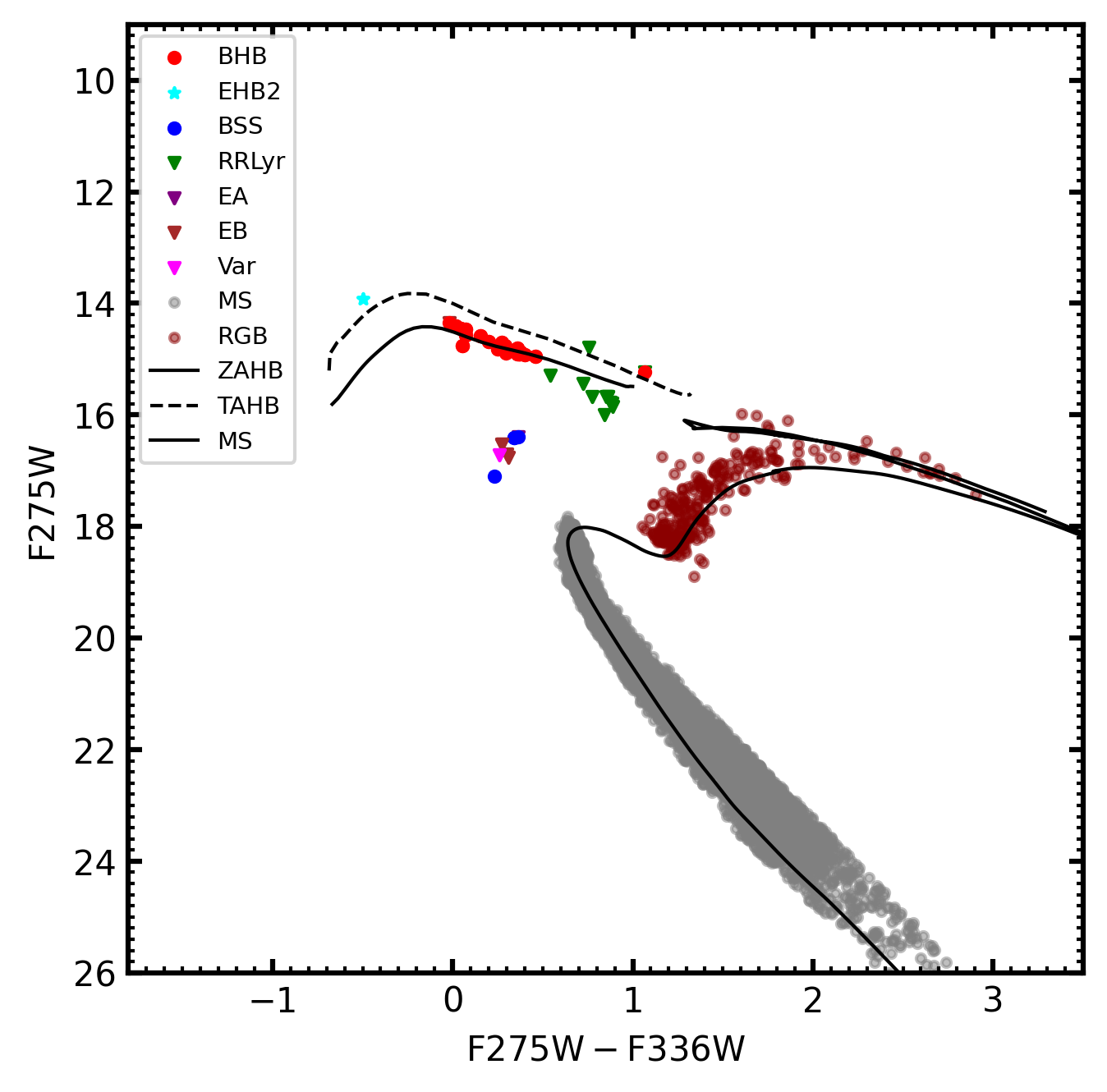} 
  \includegraphics[width=0.49\textwidth]{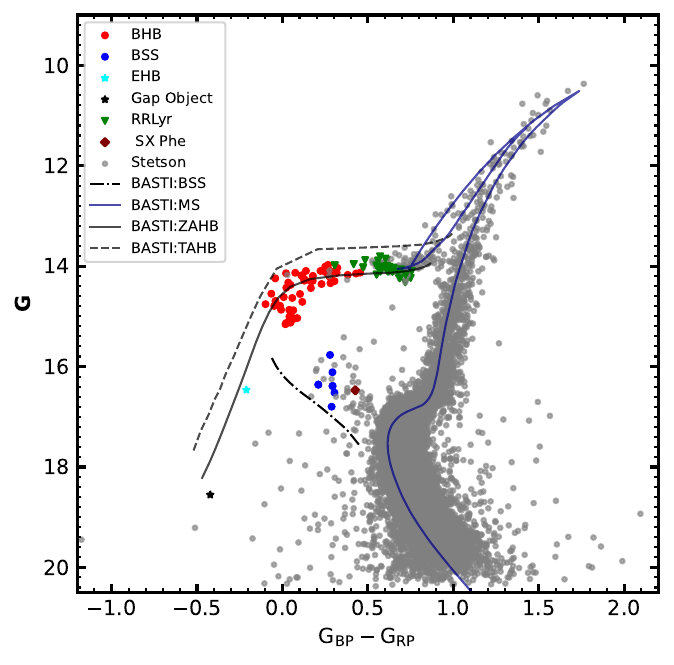}
  \includegraphics[width=0.49\textwidth]{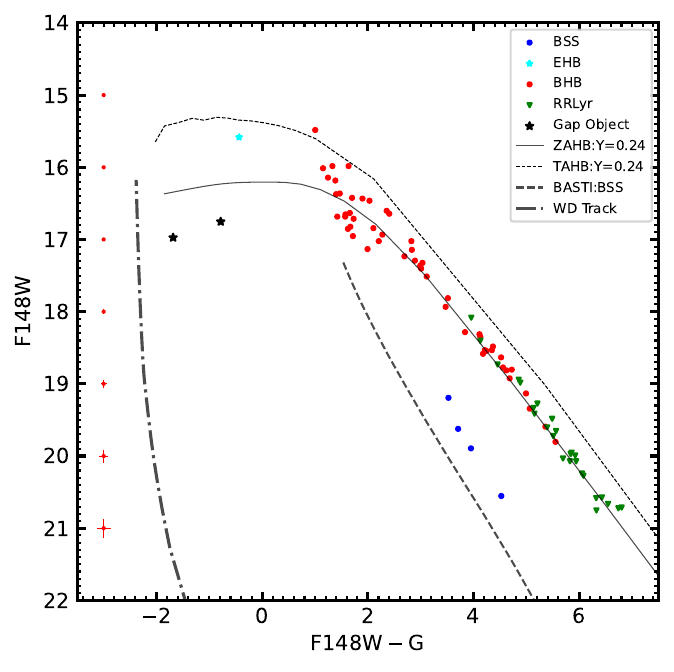}
  \includegraphics[width=0.49\textwidth]{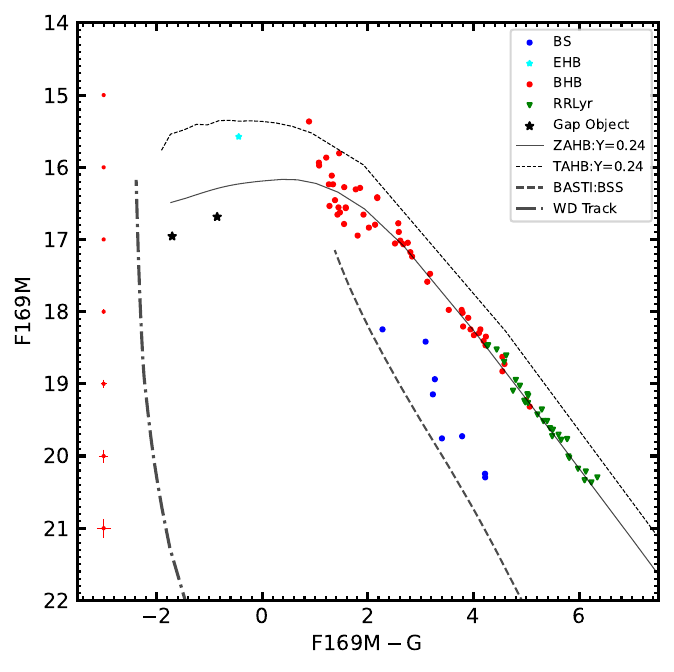}
  \caption{CMDs of GC NGC 3201. Top left panel:  F275W$-$F336W vs. F275W CMD, Top right panel: $G_{\mathrm{BP}}- G_{\mathrm{RP}}$ vs. G CMD, bottom left panel: FUV$-$optical F148W$-$G vs. F148W CMD, and bottom right panel: F169M$-$G vs. F169M CMD. Symbols in the Gaia optical CMD represent: BHBs (red filled circles), BSS (blue filled circles), EHBs (cyan stars), RR Lyrae stars (green inverted triangles), eclipsing binaries (purple/brown inverted triangles), SX Phe (brown diamonds), and Galactic foreground stars (grey/brown circles). The BASTI-IAC isochrones corresponding to an age $=$ 11.8 Gyr, [Fe/H]$= -$1.40, and DM $=$ 13.55 are overplotted on the CMDs.}
  \label{fig:CMD}
\end{figure*}

\subsection{Sources in the inner region of the cluster within the HST FOV}
We have utilized the membership probabilities from the {\it HST} UV Globular Cluster Survey (HUGS) catalog\footnote{The {\it HST} UV Globular Cluster Survey (HUGS) has observed the inner region of NGC 3201 ($\sim$3.4$' \times$ 3.4$'$).} to identify the cluster members. The stars having membership probabilities $\ge$ 60\% are considered as cluster members \citep{2016gcmp60}. We have separated 8,505 such cluster members in the inner region of the cluster which have magnitudes available in both F275W and F336W filters. To identify the evolutionary status of different cluster members, we plot these stars on F275W versus F275W$-$F336W CMD (shown in the upper left panel of Fig \ref{fig:CMD}). We also overplotted the zero-age HB (ZAHB), terminal-end HB (TAHB), and the zero-age MS (ZAMS) tracks from the Bag of Stellar Tracks and Isochrones \citep[BaSTI-IAC;][]{2018basti}. The ZAHB and TAHB tracks were generated using the age and metallicity properties of NGC 3201; however, to identify BSSs, we used the BaSTI-IAC ZAMS with age 0.5 Gyr corresponding $[\mathrm{Fe}/\mathrm{H}] = -1.40$ and mass range 1$-$1.6 $M_{\odot}$, considering main sequence turn off (MSTO) of GCs 0.8 $M_{\odot}$ \citep{Ferraro2003,Raso2017}. We have also overlaid BaSTI-IAC stellar evolutionary tracks to classify the MS and RGB stars. We followed \cite{Sahu2022} for classifying the evolved stars. The stars lying between the ZAHB and TAHB tracks are classified as HB stars. The stars lying along the ZAMS track are classified as the BSSs. We have identified 54 HB, 56 BSS, 6,202 MS and 240 RGB stars in the inner region of the cluster detected by HST.

We cross-matched the UVIT observed sources with the HST sources from the HUGS catalog using a cross-matching radius of 1.5$''$. However, with much higher spatial resolution ($< 0.1$ arcsec), almost all the UVIT sources have one of more HST counterparts. Since, UVIT only detects the hot stars such as the HB, BSS, and post-HB stars\footnote{The FUV filters of UVIT detect only hot sources of GCs, e.g., HB, post-HB, and BSSs, whereas the UV-flux of the MS and RGB stars falls below the sky background of the UV-images\citep{Piridi2026}.}, we have eliminated the MS and RGB HST stars as a counterpart to UVIT detected sources. Finally, we find 33 (35) UVIT cluster members in the inner region of NGC 3201 in the F148W (F169M) filters, of which 28 (27) and 4 (6) are HB and BSS, respectively.

\subsection{Sources of the cluster beyond HST FOV}
We have used the optical CMD using Gaia G versus $\rm G_{BP} - G_{RP}$ is shown in the upper right panel of Fig \ref{fig:CMD} for the sources outside the FOV of the {\it HST}. We followed the similar procedure to identify different types of sources in the Gaia catalog. Finally, we have identified 159 HB, 58 BSS, 17,665 MS and 627 RGB stars in the outer region of the cluster detected by Gaia.

We cross-matched the UVIT observed sources in the outer region of NGC 3201 with the Gaia EDR3 catalog of GCs \citep{gaia-gc} using a cross-matching radius of 1.5$''$. We found 85 and 94 cluster members in F148W and F169M filters, respectively. We identified 79 (83) HB stars and 4 (8) BSSs in the F148W (F169M) filters in the outer region of the cluster using the ZAHB, TAHB, and the ZAMS tracks.

\subsection{Final list of UVIT detected classified sources}
In total, we identified 23,867 MS stars, 867 RGB stars, 213 HB stars, and 114 BSSs from the combined {\it HST} and Gaia catalogs. However, we found 125 HB stars and 15 BSSs detected by UVIT in cluster NGC 3201. Among the observed HB population, we identified 19(52) BHBs and 11 (40) RR Lyrae stars in the inner (outer) region of the cluster. The remaining three HB sources were identified as variable stars: one eclipsing binary (EB) and two additional variables, using SIMBAD and the \citet{Clement2017} database. Among the UVIT detected BSS candidates, seven shows variable classification, including 1 EA (Algol-type eclipsing binary), one SX Phoenicis (SX Phe), three EB ($\beta$ Lyr–type eclipsing bin) and two as variable candidates. Excluding these variable candidates, we have performed SED on remaining 71 BHBs and eight BSSs as described in the \autoref{sec:sed fittings}. Details of the UV-bright evolved stars in the inner and outer regions are given in \autoref{app:online_tables}.

We present the UV$-$optical CMDs of the UVIT observed sources in the lower panels of Fig \ref{fig:CMD}. The ZAHB, TAHB, and ZAMS tracks of BaSTI-IAC are overlaid on the CMDs. The WD track adopted from \citet{Bergeron2005} with 0.5 $M_{\odot}$ \citep{Kalirai_2009} is also overplotted on both the CMDs. The observed sources align well with the theoretical tracks in both CMDs. 

The UV-optical CMDs reveal that the outer BHB population spans a wide range in magnitude and color: 15.5$-$19.8 mag in F148W, 15.3$-$19.3 mag in F169M, 1.1$-$5.6 mag in F148W$-$G, 1.0$-$5.1 mag in F169M$-$G. In contrast, the inner BHB population (from HST data) spans 0.12$-$1.2 in F275W$-$F336W, while the outer BHB population exhibits a color spread of $\rm -0.1 \ mag < G_{BP} - G_{RP} < $ 0.45 mag with the G band magnitude ranging from 14.13$-$14.75 mag. Even after accounting for differential reddening, \citet{Kravtsov2009} reported significant broadening in the horizontal branch morphology of NGC 3201 across multiple optical CMDs (B$-$V vs. V, V$-$I vs. I, and B$-$I vs. U). Such broadening is commonly attributed to variations in mass loss along the red giant branch (RGB) tip \citep{D'Antona2002, catelan2009horizontalbranchstarsinterplay, Gratton2012, Tailo2020MNRAS}. The inner BSS candidates identified from the {\it HST} catalog span a  0.4$-$0.5 mag color range in F275W$-$F336W in the top-left {\it HST} CMD Fig~\ref{fig:CMD}, and the outer region BSS  occupy a color range 0.2$-$0.4 mag in $\rm G_{BP}-G_{RP}$. The BSS population spans $\sim$1.55 mag in F148W and $\sim$1.22 mag in F169M.

In CMDs, we identify two gap objects (GOs) among the post-HB stars. The GOs appear bright and extremely blue with F148W$-$G color of $\sim -$1.6 and $\sim -$0.7 mag, respectively. Their locations in the CMD place them beyond the He-core burning phase and above the WD cooling tracks, suggesting that they occupy a transitional evolutionary stage and are therefore classified as gap objects. For GO2, which lacks $\rm G_{BP} - G_{RP}$ photometry, we identified the most reliable Gaia counterpart via cross-matching with the Stetson catalog \citep{Stetson2019}, choosing the hotter (CMD-consistent) source. Although \cite{Haurberg2010} found that GOs in M15 can include CVs, cross-matching with existing catalog of CVs and X-ray sources in NGC 3201 \citep{Downes2001, Webb2006} revealed no matches.

The BSS were selected based on their position in the FUV CMDs, just below the HB sequence (bottom panels of Fig \ref{fig:CMD}). The selection was cross-checked using optical CMD to ensure consistency with the BS sequence.  The population of BSS spans $\sim$2 mag in F148W  and $\sim$2.3 mag in F169M. In the optical CMDs, they occupy a color range 0.2 $-$ 0.5 mag in $\rm G_{BP}-G_{RP}$ and a magnitude spread of $\sim$1.7 mag in G band.

We classify EHB stars among the remaining post-HB population primarily by their location at the extreme blue end of the HB in the CMDs (Fig.~\ref {fig:CMD}) and by their significantly higher effective temperatures compared to BHB stars. EHB1 shows color of $-$0.35 in F275W$-$F336W, while EHB2 has a color of $-$0.33 in F148W$-$G.  These stars appear very bright in the UV while remaining relatively faint in optical bands, consistent with their high effective temperatures and luminosities.

\section{SED fitting and estimation of physical parameters of UV-bright stars} \label{sec:sed fittings}

\begin{figure*}[h!]
  \centering
  \includegraphics[width=0.49\textwidth]{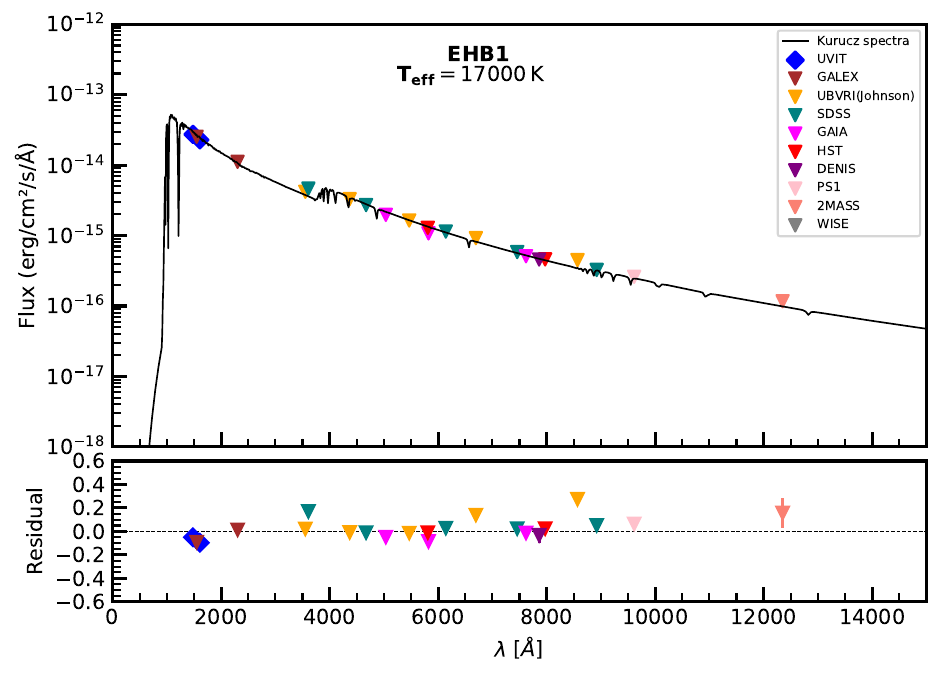}
  \includegraphics[width=0.49\textwidth]{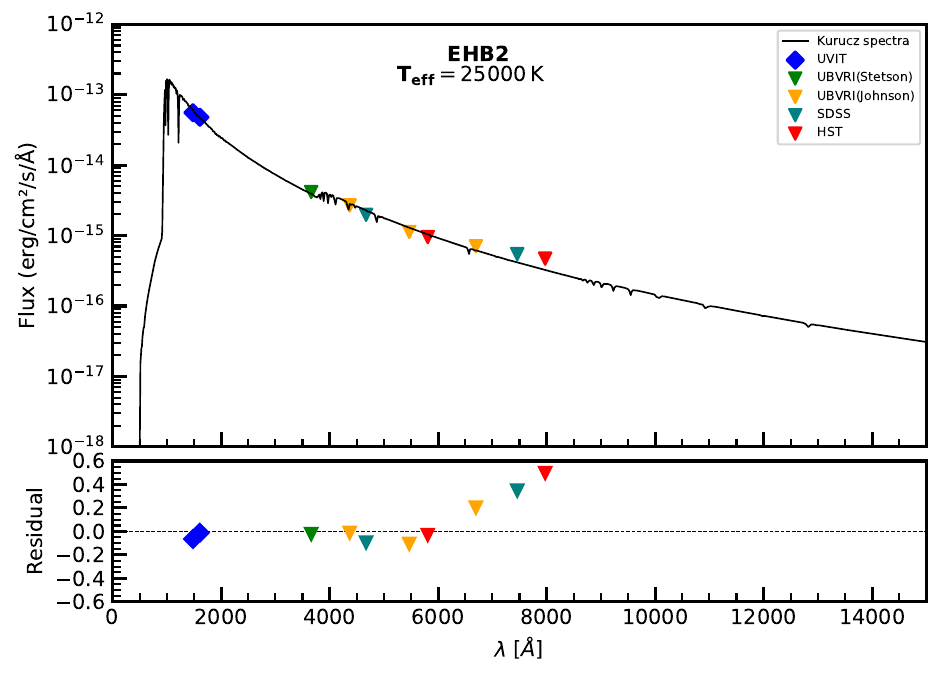}
  \includegraphics[width=0.49\textwidth]{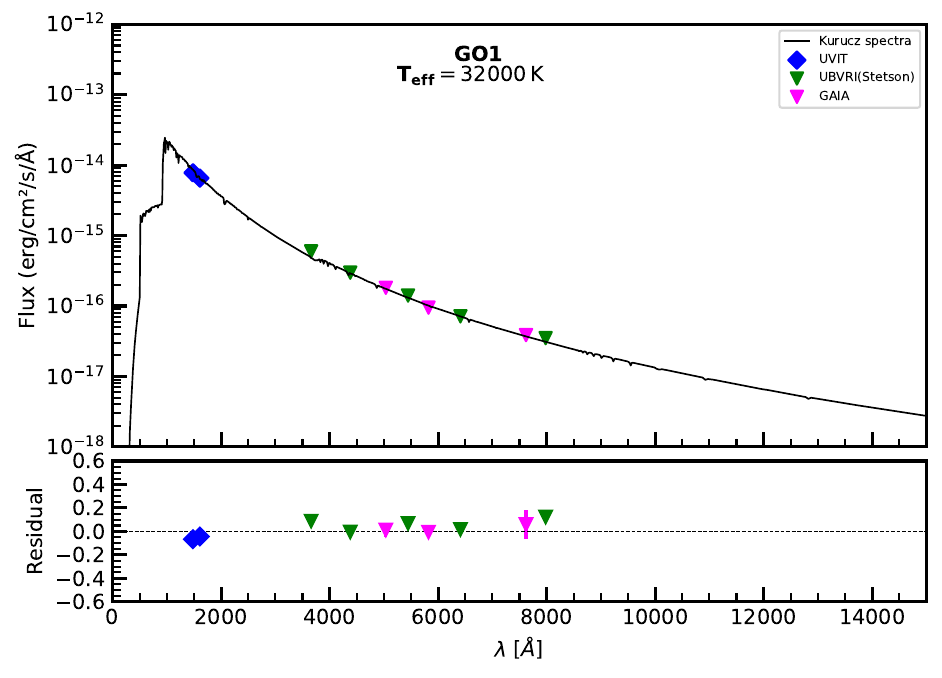}
  \includegraphics[width=0.49\textwidth]{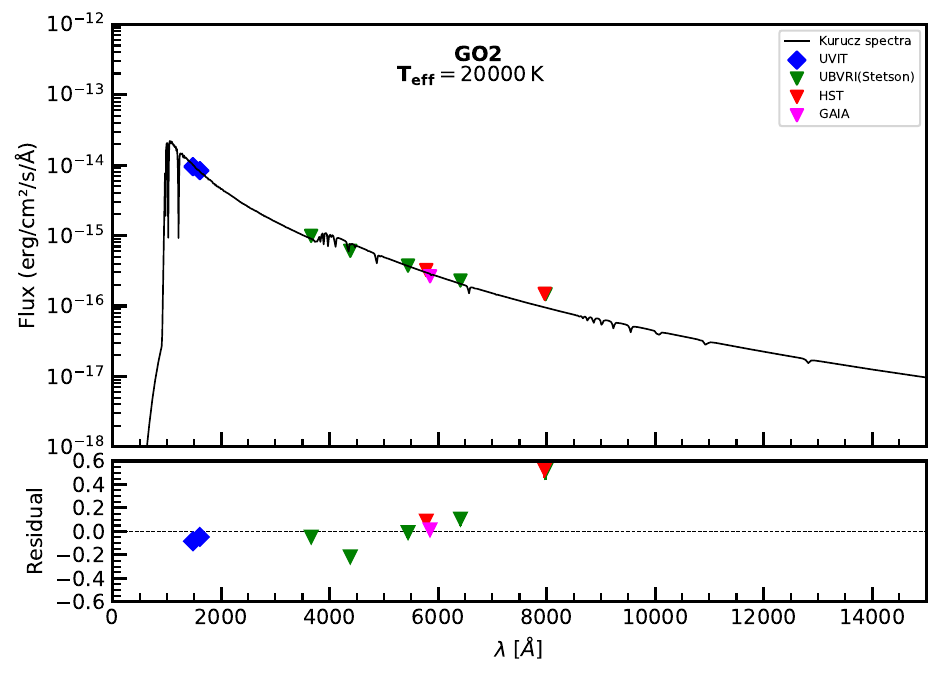} 
  \caption{Spectral energy distribution of EHBs and GO objects. In legend, blue diamonds are UVIT fluxes, and color triangles for other bands. The Kurucz model with best-fit parameters is overplotted on the residuals of the respective SEDs in the lower panel.}
  \label{fig:SED}
\end{figure*}

\begin{figure}
    \centering
    \includegraphics[width=\columnwidth]{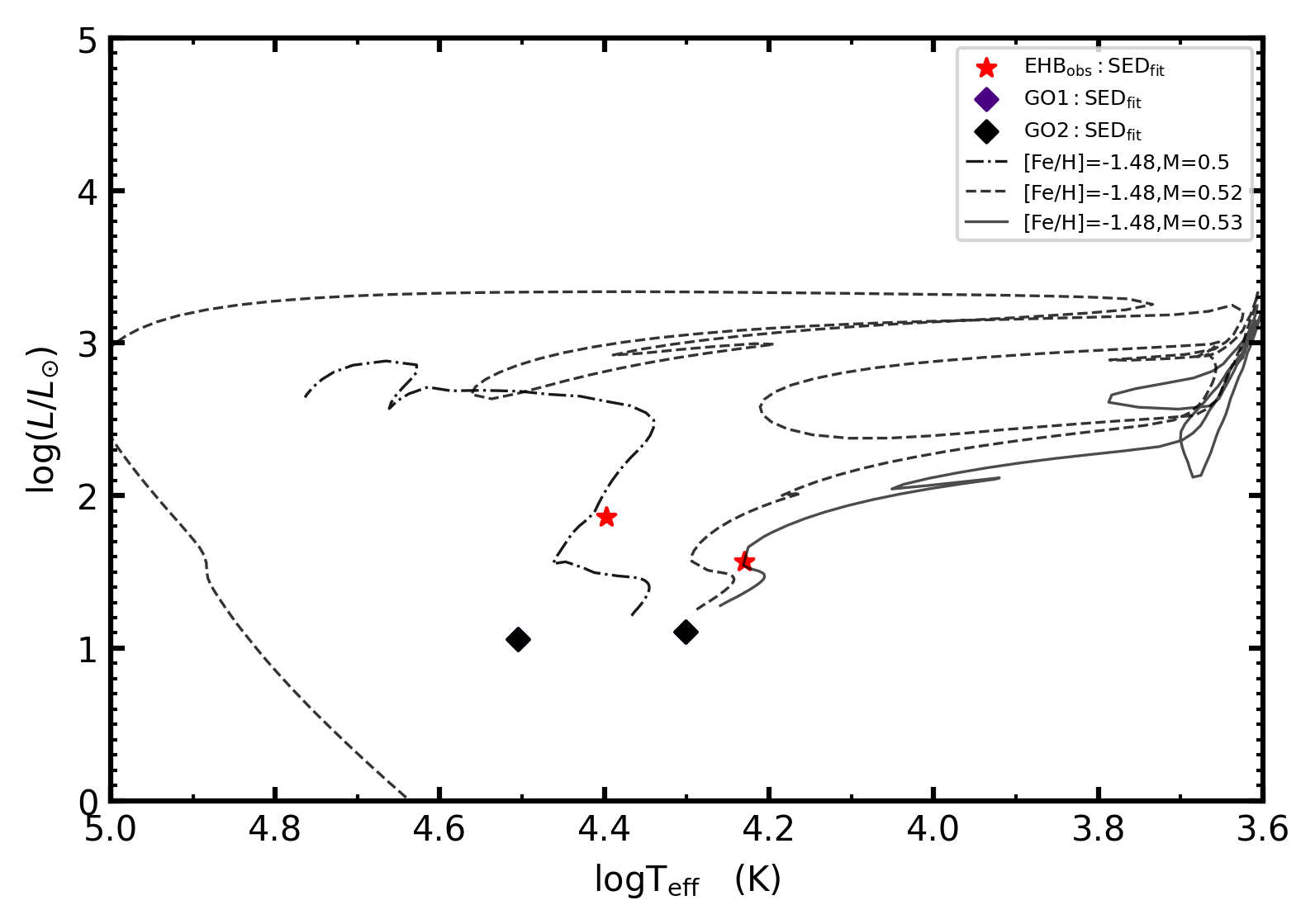}
    \caption{Post-HB evolutionary models with a metallicity of $-$1.48 and a mass range of $0.48-0.54  M_{\odot}$ \citep{Dorman1993}. Red stars present two EHB stars, and color diamonds are gap objects. The log $\rm T_{eff}$ and log(L/$\rm L_{\odot}$) are obtained from SED fit results.}
    \label{fig:EHB_tracks}
\end{figure}

\begin{figure}
    \centering
    \includegraphics[width=\columnwidth]{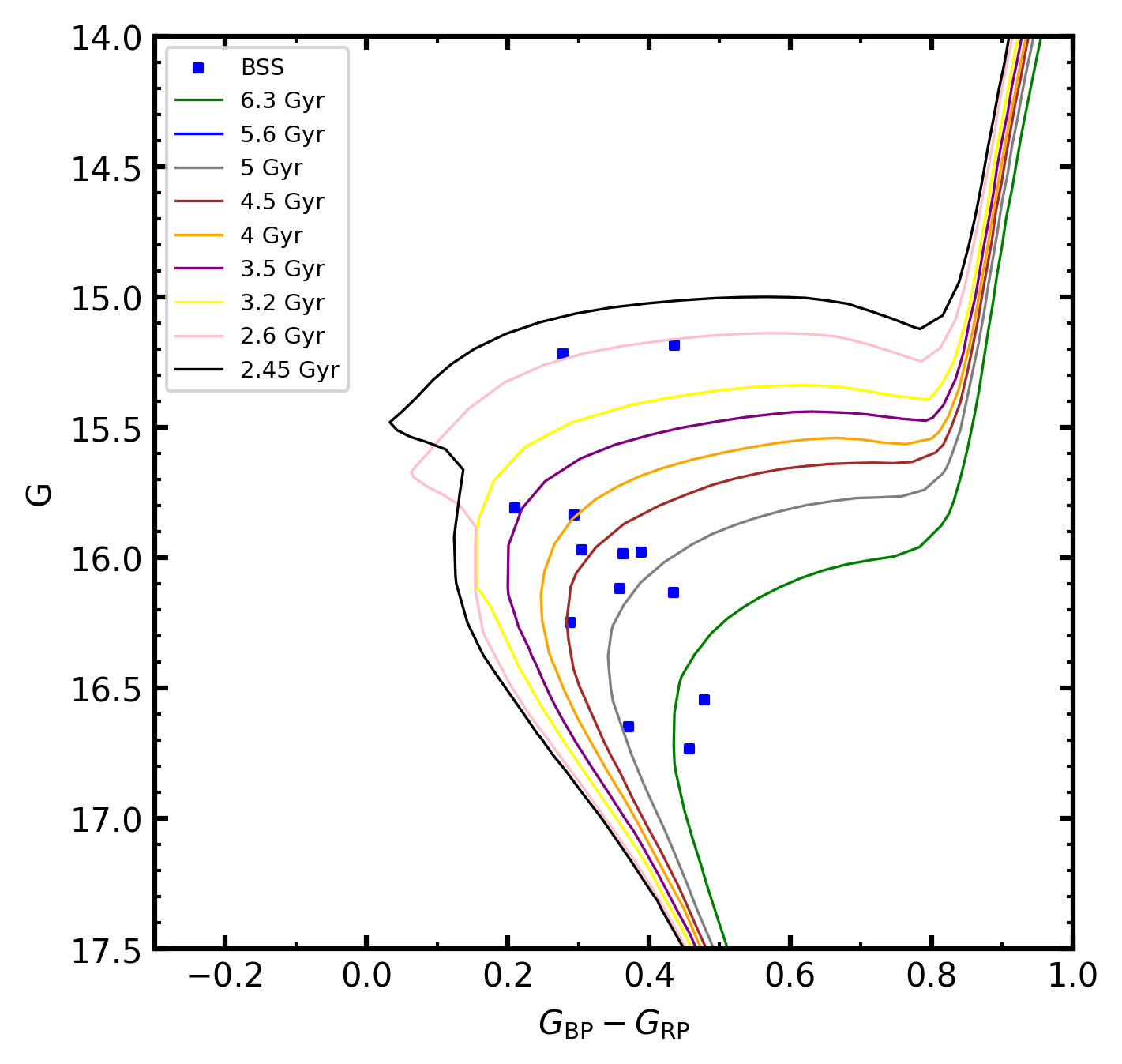}
    \caption{Identified BSSs (blue squares) are shown in $G_{\mathrm{BP}} - G_{\mathrm{RP}}$ vs. G CMD  are fitted with PARSEC \citep{Bressan2012} isochrones with ages 2.45 to 6.3 Gyr and [Fe/H] = $-$1.5.} 
    \label{fig:BSS_iso}
\end{figure}

We determined physical parameters of UV-bright stars including effective temperature ($\rm T_{eff}$), luminosity ({$\rm L_{bol}$}), surface gravity ({$\rm log\ g$}), radius, and mass using spectral energy distribution (SED) fits in the Virtual Observatory SED analyzer (VOSA; \cite{VOSA}). We used photometric data of UVIT (F148W and F169M), {\it GALEX} \citep[FUV, NUV;][]{Bianchi2014}, Gaia \cite[G, $\rm G_{BP}$, $\rm G_{RP}$;][]{GaiaCollaboration}, SDSS \citep[u, g, r, i, z;][]{SDSS}, Johnson \citep[UBVRI;][]{1968jhonson_ubvri}, {\it HST} \citep[F275W, F336W, F438W, F606W, F814W ;][]{Sirianni2005,Marinelli2025}, 2MASS \citep[J, H, Ks;][]{Skrutskie2006}, DENIS \citep[J, H, Ks;][]{Fouque2000}, and WISE \citep[W1, W2, W3, W4;][]{wright2010} archival catalogs and fitted the Kurucz stellar atmospheric model \citep{Castelli1997,Kurucz2003} to the observed SED fluxes of the UV-bright sources. SED analysis was restricted to only non-variable stars. 

\textbf{GO objects:} We found that the $\rm T_{eff}$ of the GOs are best fit by 32,000 K and 20,000 K. Based on their locations between the post-EHB and WD evolutionary tracks,  and their derived properties, these objects are likely the BHk candidates \citep{Dieball2009, Brown2016}. They appear to form a hook-like structure at the blue end of the horizontal branch \citep{D'Cruz1996}.  However, spectroscopic data are necessary to confirm their precise atmospheric parameters and evolutionary status. The SED fitted parameters of the GOs are summarized in Table~\ref{tab: EHB_SED_TABLE}, with the corresponding SED fits shown in Fig~\ref{fig:SED}. 

\textbf{EHBs:} We fitted the observed fluxes of the two identified EHBs with Kurucz models, adopting a metallicity range of $-$2 to $-$1 and log g = 5, assuming sub-dwarf nature. The SED fit for both EHBs is shown in the Fig~\ref{fig:SED}. The high $\chi^2_{\mathrm{red}}$ values, despite an apparent good fit, are due to small observational flux errors. For this, VOSA has introduced Vgfb to estimate goodness of fit ($Vgfb < 15$ indicates a good fit). Moreover, the final SED solutions were selected not only in terms of minimum $\chi^2_{\mathrm{red}}$  values, but also in terms of overall agreement of the model with the observed photometric data over the entire range of wavelengths. The EHB1 shows a temperature of 17000 K, placing it on the cooler end of the EHB.  Its derived radius $\sim 0.703\ R_{\odot}$ and luminosity $\sim 37.272\ L_{\odot} $ suggest that it is slightly more massive and is not yet fully contracted. The second EHB shows an effective temperature of 25000 K with luminosity $\sim$ 73.971 $L_{\odot} $ and radius $\sim$ 0.458 $R_{\odot}$, suggesting it lies within the typical EHB range. The SED-based estimates are listed in the Table~\ref{tab: EHB_SED_TABLE}. To estimate the physical parameters like mass, age, surface gravity, envelope mass, core temperature, and density of EHB, we used evolutionary models of evolved stars. We adopted tracks with a metallicity of $-$1.48 and a mass range of $0.48-0.54  M_{\odot}$ from \citep{Dorman1993, 1994yCat.6065....0D} theoretical models of post-HB evolution. We found that EHB1 and EHB2 lie close to the evolutionary track with total masses of 0.53 and $0.5 M_{\odot}$,  with corresponding envelope masses 0.0448 M$_{\odot}$ and 0.0148 M$_{\odot}$, respectively, as seen in Fig~\ref{fig:EHB_tracks}. The derived parameters are given in the Table~\ref{tab:EHB_tracks}. 

\textbf{BSSs:} The fundamental parameters of the observed UV-bright BSSs such as $\rm T_{eff}$, Luminosity, and Radius, were derived from the SED fit (listed in the Table~\ref{tab:BSS TABLE}). The BSSs show $T_{eff}$ in the range 7500$-$8500 K, $L / L_{\odot}$ in the range 3.9$-$17.8, and $R / R_{\odot}$ spanning 0.86 to 1.94. We used isochrones from  PARSEC version 1.2S \citep{Bressan2012}, OBC \citep{Girardi2002, Marigo2017} library of ages ranging from 2.69 Gyr to 5.62 Gyr and metallicity $-$1.5 to fit the BSSs. By comparing BSSs positions on the Gaia CMD to the nearest point on the theoretical isochrones (Fig~\ref{fig:BSS_iso}), we obtained isochrone-inferred masses and the BSSs apparent isochrone age, which are listed in Table~\ref{tab:BSS TABLE}. It is important to note that the apparent age of the isochrones cannot be taken as a real physical age, since BSSs are rejuvenated stars formed through mass transfer in binary systems or through stellar collisions. \citep{McCrea1964, Ferraro2009, Ferraro:2012dg} The observed masses of BSSs were found to be less than 2 $M_{\odot}$. 
   
\textbf{GO objects:} We found that the $\rm T_{eff}$ of the GOs are best fit by 32,000 K and 20,000 K. Based on their locations between the post-EHB and WD evolutionary tracks,  and their derived properties, these objects are likely the BHk candidates \citep{Dieball2009, Brown2016}. They appear to form a hook-like structure at the blue end of the horizontal branch \citep{D'Cruz1996}.  However, spectroscopic data are necessary to confirm their precise atmospheric parameters and evolutionary status. The SED fitted parameters of the GOs are summarized in Table~\ref{tab: EHB_SED_TABLE}, with the corresponding SED fits shown in Fig~\ref{fig:SED}. 

\textbf{EHBs:} We fitted the observed fluxes of the two identified EHBs with Kurucz models, adopting a metallicity range of $-$2 to $-$1 and log g = 5, assuming sub-dwarf nature. The SED fit for both EHBs is shown in the Fig~\ref{fig:SED}. The high $\chi^2_{\mathrm{red}}$ values, despite an apparent good fit, are due to small observational flux errors. For this, VOSA has introduced Vgfb to estimate goodness of fit ($Vgfb < 15$ indicates a good fit). Moreover, the final SED solutions were selected not only in terms of minimum $\chi^2_{\mathrm{red}}$  values, but also in terms of overall agreement of the model with the observed photometric data over the entire range of wavelengths. The EHB1 shows a temperature of 17000 K, placing it on the cooler end of the EHB.  Its derived radius $\sim 0.703\ R_{\odot}$ and luminosity $\sim 37.272\ L_{\odot} $ suggest that it is slightly more massive and is not yet fully contracted. The second EHB shows an effective temperature of 25000 K with luminosity $\sim$ 73.971 $L_{\odot} $ and radius $\sim$ 0.458 $R_{\odot}$, suggesting it lies within the typical EHB range. The SED-based estimates are listed in the Table~\ref{tab: EHB_SED_TABLE}. To estimate the physical parameters like mass, age, surface gravity, envelope mass, core temperature, and density of EHB, we used evolutionary models of evolved stars. We adopted tracks with a metallicity of $-$1.48 and a mass range of $0.48-0.54  M_{\odot}$ from \citep{Dorman1993, 1994yCat.6065....0D} theoretical models of post-HB evolution. We found that EHB1 and EHB2 lie close to the evolutionary track with total masses of 0.53 and $0.5 M_{\odot}$,  with corresponding envelope masses 0.0448 M$_{\odot}$ and 0.0148 M$_{\odot}$, respectively, as seen in Fig~\ref{fig:EHB_tracks}. The derived parameters are given in the Table~\ref{tab:EHB_tracks}. 

\textbf{BSSs:} The fundamental parameters of the observed UV-bright BSSs such as $\rm T_{eff}$, Luminosity, and Radius, were derived from the SED fit (listed in the Table~\ref{tab:BSS TABLE}). The BSSs show $T_{eff}$ in the range 7500$-$8500 K, $L / L_{\odot}$ in the range 3.9$-$17.8, and $R / R_{\odot}$ spanning 0.86 to 1.94. We used isochrones from  PARSEC version 1.2S \citep{Bressan2012}, OBC \citep{Girardi2002, Marigo2017} library of ages ranging from 2.69 Gyr to 5.62 Gyr and metallicity $-$1.5 to fit the BSSs. By comparing BSSs positions on the Gaia CMD to the nearest point on the theoretical isochrones (Fig~\ref{fig:BSS_iso}), we obtained isochrone-inferred masses and the BSSs apparent isochrone age, which are listed in Table~\ref{tab:BSS TABLE}. It is important to note that the apparent age of the isochrones cannot be taken as a real physical age, since BSSs are rejuvenated stars formed through mass transfer in binary systems or through stellar collisions. \citep{McCrea1964, Ferraro2009, Ferraro:2012dg} The observed masses of BSSs were found to be less than 2 $M_{\odot}$. 
   
We found that the BHBs in NGC 3201 have a $\rm T_{eff}$ range from 7250 to 11250 K, a luminosity between 22 and 64  $L_{\odot}$, and a radius spanning 1.5 to 4.25 $R_{\odot}$. These BHBs become progressivel more compact as they move from the cooler to the hotter end of the sequence, consistent with expected evolutionary behavior along ZAHB. The detailed estimated stellar parameters of the individual BHBs are summarized in the Table~\ref{tab:HB-SED-table}. 

\begin{table*}
\centering
\caption{Physical parameters of EHBs and GOs derived from SED fittings}
\label{tab:EHB_tracks}
\smallskip
\resizebox{\textwidth}{!}{
    \begin{tabular}{l r r l r r l l r r r} 
    \toprule
    Obj.ID & RA & DEC & $T_{\rm eff}$ (K) & $\log g$ & [Fe/H] & $R/R_{\odot}$ & $L/L_{\odot}$ \\
    \midrule
      EHB1 & $154.2821$ & $-46.5394$ & $17000 \pm 500$ & 5 & $-1.0$ & $0.703 \pm 0.041$ & $37.272 \pm0.677$ \\
      EHB2 & $154.4066$ & $-46.4186$ & $25000 \pm 500$ & 5 & $-1.5$ & $0.458 \pm 0.018$ & $73.971\pm 0.225$ \\
      GO1 & $154.4090$ & $-46.3732$ & $32000 \pm 500$ & 4 & $-1.0$ & $0.11 \pm 0.003$ & $11.48 \pm 0.017$\\
      GO2 & $154.4310$ & $-46.4416$ & $20000 \pm 500$ & 5 & $-1.5$ & $0.292 \pm 0.014$ & $12.850 \pm 0.121$\\
    \bottomrule
    \end{tabular}
}
\end{table*}

\begin{table*}
 \centering
 \caption{Atmospheric parameters of EHBs obtained from evolutionary tracks} 
 \label{tab: EHB_SED_TABLE}
   \begin{tabular}{|l|r|r|r|r|r|r|r|}
   \hline
      \multicolumn{1}{|c|}{Obj.ID} &
      \multicolumn{1}{c|}{YHB} &
      \multicolumn{1}{c|}{[O/Fe]} &
      \multicolumn{1}{c|}{$M/M_{\odot}$} &
      \multicolumn{1}{c|}{Age (Myr)} &
      \multicolumn{1}{c|}{Msh} &
      \multicolumn{1}{c|}{logTc} &
      \multicolumn{1}{c|}{log$\rho_{c}$} \\
    \hline
      EHB1 & 0.247 & 0.6 & 0.53 & 102.6 & 0.4852 & 8.2161 & 4.5048\\
      EHB2 & 0.247 & 0.6 & 0.5 & 115.8 & 0.4852 & 8.1386 & 5.3072\\
    \hline
    \end{tabular} 

{
 Note: YHB represents He composition, Age (Myr) since ZAHB, Msh is Mass at peak energy production rate of hydrogen burning shell, $\log T_{\rm c}$ and $\log \rho_{\rm c}$ are central temperature (K) and central density in cgs units, respectively.
 \par}  
\end{table*}

\begin{table*}
 \centering
 \caption{Summary of physical parameters of eight BSSs derived from SED fitting and evolutionary tracks}  
 \label{tab:BSS TABLE}
     \resizebox{\textwidth}{!}{
    \begin{tabular}{l r r l l l r r r r r} 
    \toprule
    Obj.ID &RA &DEC& $T_{\rm eff}$ (K) & $L/L_{\odot}$ & $R/R_{\odot}$ & Age (Gyr) & $M/M_{\odot}$ & $\log g$\\
    \midrule
    \noalign{\vskip 2pt}
     BSS1 & $154.3994$ & $-46.3927$ & $8750\pm 125$ & $ 3.964\pm 0.005$ & $ 0.866\pm 0.024$ & -- & -- & 4 \\
     BSS2 & $154.3786$ & $-46.3333$ & $ 7500\pm 125$ & $8.256\pm0.003 $ & $1.701 \pm0.056$ & 3.98 & 1.086  & 4.5\\
     BSS3 & $154.4331$ & $-46.4138$ & $8500 \pm 125$ & $8.718 \pm 0.004 $ & $1.370 \pm 0.040 $ & 4.79 & 1.028 & 4.5\\
     BSS4 & $154.4373$ & $-46.2439$ & $8250 \pm 125$ & $9.400 \pm 0.002$ & $1.500 \pm 0.045$ & 3.98 & 1.086 & 4.0\\
     BSS5 & $154.3862$ & $-46.4501$ & $8500 \pm 125$ & $10.200 \pm 0.003$ & $1.472\pm 0.043$ & 3.55 & 1.120 & 4.5\\
     BSS6 & $154.3610$ & $-46.3882$ & $8500 \pm 125$  & $17.828 \pm 0.005$ & $1.947  \pm 0.057$ & 2.69 & 1.237 & 4.0\\
     BSS7 & $154.4034$ & $-46.4137$ & $8000 \pm 125$ & $7.920 \pm 0.004$ & $1.465 \pm 0.044$ & 5.62 & 0.985 & 5.0\\
     BSS8 & $154.3609$ & $-46.3986$ & $8500 \pm 125$ & $7.006  \pm 0.004$ & $1.220  \pm 0.035$ & 4.37 & 1.040 & 5.0\\
    
    \hline\end{tabular}
 }
\end{table*}

\begin{table*}
 \centering
 \caption{UV magnitudes and best-fit physical parameters of the UV-bright BHB stars. The complete table is available online.} 
 \label{tab:HB-SED-table}
 \resizebox{\textwidth}{!}{
    \begin{tabular}{l l l l l l l l l l l} 
    \toprule
    Obj.ID & RA & DEC & F148W & $\rm \sigma_{F148W}$ & F169M & $\rm \sigma_{F169M}$ & $T_{\rm eff}$ (K) & $\log g$ & $R/R_{\odot}$ & $L/L_{\odot}$ \\
    \midrule
    BHB1 & 154.6348 & $-46.4270$ & 19.804 & 0.12 & 19.316 & 0.12 & 7750 & 3.5 & 3.621  $\pm$ 0.116 & $42.622 \pm 0.017$ \\
    BHB2 & 154.4143 & $-46.4055$ & 19.354 & 0.08 & 18.906 & 0.09 & 7250 & 3.0 & 3.745  $\pm$ 0.121 & $46.356 \pm 0.006$ \\
    BHB3 & 154.4771 & $-46.3974$ & 19.344 & 0.09 & 18.826 & 0.10 & 7750 & 3.5 & 3.640  $\pm$ 0.117 & $43.070 \pm 0.011$  \\
    $-$ & $-$ & $-$ & $-$ & $-$ & $-$ & $-$ & $-$ & $-$ & $-$ & $-$ \\
    \hline
    \end{tabular}
  }
\end{table*}

\section{Radial Distribution and Cluster Dynamics} \label{sec:Radial distribution}

\begin{figure}
    \centering
    \includegraphics[width=\columnwidth]{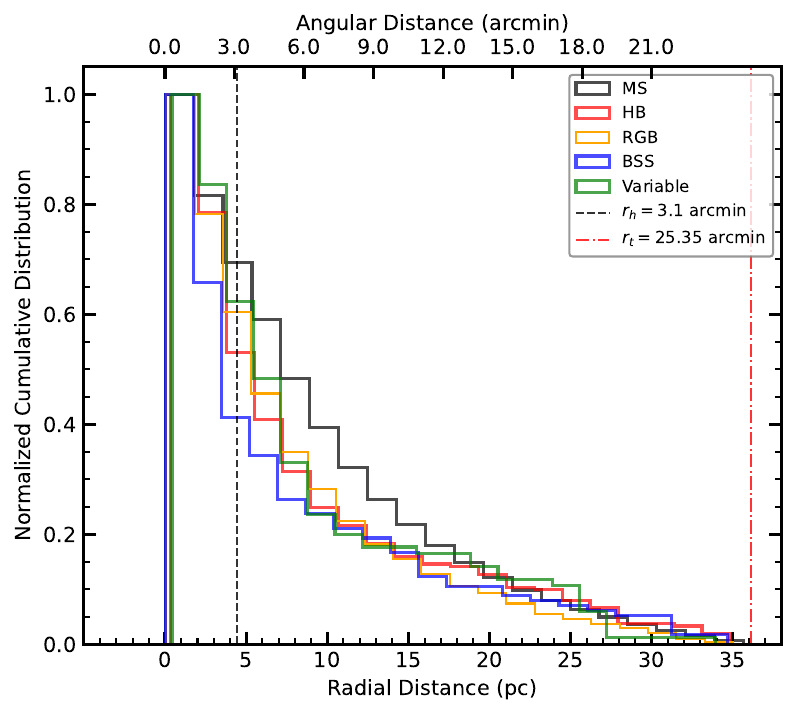}
    \caption{Normalized cumulative distribution for MS, HB, RGB, BSS and Variable stars. The half-light radius and tidal radius are plotted as dashed lines.}
    \label{fig:radial dist.}
\end{figure}

Radial distribution provides key insights into cluster dynamics and helps constrain the formation mechanisms of different stellar populations. We have constructed the normalized cumulative distribution (NCD) for the complete stellar population from {\it HST}-Gaia detections, with UVIT detections representing a subset of these sources where available: MS, HB, RGB, BSS, and variable stars. As shown in Fig ~\ref{fig:radial dist.}, all populations show centrally concentrated radial distributions, with a gradual flattening well within the cluster's tidal radius \citep[$r_{\mathrm{t}}$ = 25.35$'$; ][]{harris2010newcatalogglobularclusters}. The HB population has a slightly steeper radial distribution compared to MS stars. However, with a relatively smaller number of HBs in comparison with the MS population in the outer bins, the result can be due to a combination of weak mass segregation and statistical effects. All the BHBs are distributed within 20 pc of the centre.  Of these, 40 BHBs lie inside the half-light radius \citep[$r_{\mathrm{h}} = 3.1'$;][]{Harris1996}, while 31 lie outside it. Only 20\% of the BHBs lie within the core radius. The variable stars show an NCD that lies between those of the BSS and MS stars, with their distribution extending up to 33 pc of the cluster's centre. Among the EHBs and GOs, EHB1, GO1, and GO2 lie outside the core radius \citep[$r_{\mathrm{c}} =  1.3'$;][]{Harris1996} at projected radial distances of $ 9.117 ^{\prime}, 2.368 ^{\prime}$ and $ 2.087^{\prime}$. Only EHB2, at $0.391^{\prime}$, is located within the core radius.  
  
\begin{figure}
    \centering
    \includegraphics[width=\columnwidth]{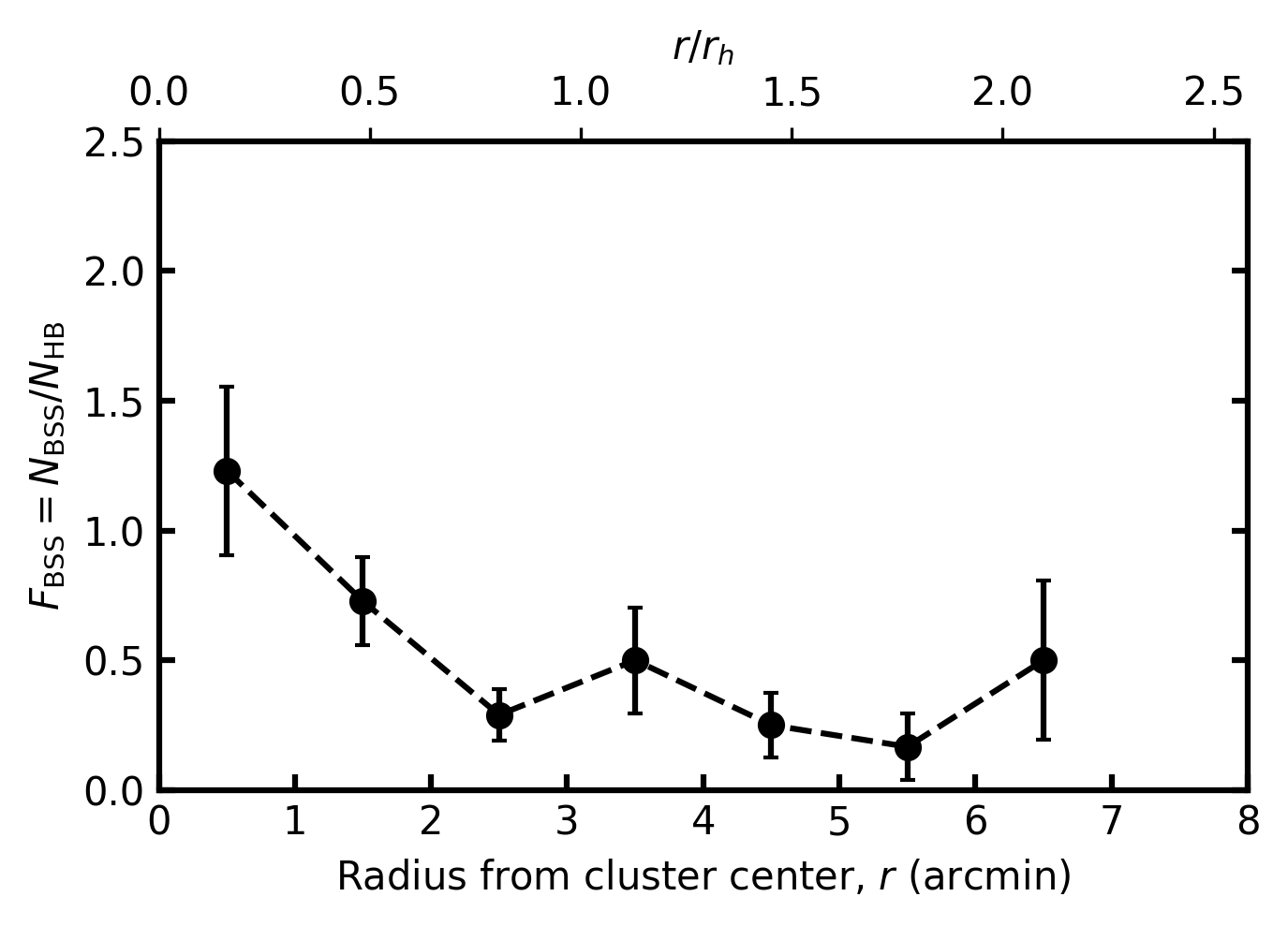}
    \caption{Specific frequency plot for BSS distribution with respect to HB population. The bottom axis is radial distance from the centre, and the top x-axis represents the scaled $r/r_{\mathrm{h}}$, where $r_{\mathrm{h}}$ is the half-light radius.}
    \label{fig:specific frequency}
\end{figure}

In contrast, the BSSs show a sharp, steep rise in the NCD within the inner 5 pc, highlighting their strong central concentration. 60\% BSSs lie inside the half-light radius, and 36\% lie inside the core radius. The radial distribution of BSSs provides important insights into the cluster's dynamical age \citep{Ferraro:2012dg}. We have calculated the specific frequency of BSSs ($  F_{\rm BSS}  $) with respect to the HB population. We divided the observed field into concentric annuli of 1$'$ width and counted the number of BSSs and HB stars in each annulus. To construct a homogeneous sample and improve the statistical robustness, we selected HB and BSS populations over the combined coverage of {\it HST} in inner region and Gaia in the outer region, including additional BSSs reported in the literature \citep{Simunovic2014, Simunovic2016}, resulting in a final sample of 120 BSSs and 213 HB cluster members. The radial distribution of specific frequency represents the spatial variation of BSSs within the cluster. The radial profile of the specific frequency is shown in the Fig~\ref{fig:specific frequency}, plotted against the projected distance from the cluster center, with $r/ r_{\mathrm{h}}$ also indicated on the upper X-axis. Poisson errors are shown for each bin centered at every 0.5$'$ intervals. The specific frequency of BSSs in NGC 3201 shows central enhancement and generally decreases with increasing radius; the apparent increase in the outermost bin is due to small numbers of stars and not a statistically significant feature. As reported in several globular clusters, such as M3 and 47 Tuc, bimodality refers to a statistically significant central peak followed by an intermediate-radius minimum and an outer rising branch, associated with an intermediate dynamical cluster age \citep{2004ferraro,ferraro2012}. This absence of bimodality indicates that NGC 3201 is dynamically evolved, but has not yet reached a fully relaxed state or undergone core collapse.

\section{Summary and Conclusion} \label{sec:conclusion}
We present a comprehensive study of UV-bright stellar populations in the globular cluster NGC 3201 using FUV observations from the UVIT onboard Astrosat. We combined UVIT photometry in the F148W (1481 \AA) and F169M (1608 \AA) filters with optical data from HST and Gaia to construct UV-optical CMDs and using the location of the objects on CMDs, we classified the detected sources into various evolutionary stages. We found 71 BHBs, 15 BSS, and 51 RR Lyrae detected by UVIT. In addition, we also identified two EHBs and two bright GOs (possibly BHk candidates). Stellar evolution models from the BASTI-IAC database, computed for an age of 11.80 Gyr and [Fe/H] $= -$1.40, fits well to the optical CMD. The fit yields a DM of 13.55 $\pm$ 0.05 mag, corresponding to a distance of 5.13 $\pm$ 0.12 kpc. 

Using the SED analysis, we determined the physical parameters $\rm T_{eff}$, luminosity, and radius of detected hot UV-bright stars. We found that BHB stars exhibit effective temperatures ranging from 7250 K to 11250 K. In the analysis, we identified two EHBs: one located inside the cluster core radius and the other outside with temperatures  $\rm T_{eff} =$ 17,000 K and 25,000 K respectively. Using post-HB evolutionary model from \citep{Dorman1993}, we estimate their masses to be 0.53 and 0.5 $\rm M_{\odot}$, respectively. UV CMDs also enabled the detection of two bright gap objects (possibly BHks) in the post-HB phase lying within the half-light core of the cluster, with derived effective temperatures of 32,000 and 20,000 K  respectively. However, spectroscopic data are necessary to confirm the exact atmospheric parameters and evolutionary pathway of these objects. For the eight non-variable BSSs, the derived parameters obtained from the SED fittings lie in the range $  T_{\rm eff} = 7,500{-}8,500  $ K, radius 0.86 to 1.94 $\rm R_{\odot}$, Masses 0.985 $-$ 1.242 $\rm M_{\odot}$. These masses are significantly higher than those of normal main-sequence turn-off stars, supporting their blue straggler classification. The isochrone ages BSSs are found to be 2.7-5.6 Gyr.

To understand the dynamical evolution of NGC 3201, we examined the radial distribution of different stellar populations. We found the BHB, MS, RGB and varibles are uniformly distributed throughout the cluster and have centrally concentrated radial distributions. The BSSs exhibit the strongest central concentration, as revealed by both its normalized cumulative radial distribution and its specific frequency relative to the HB population. Thus, results show consistency with old and dynamically evolved clusters' mass segregation with population-based radial distribution evidence.

\section*{Acknowledgments}
This publication uses the data from the {\it AstroSat} mission of the ISRO, archived at the Indian Space Science Data Center (ISSDC). ACP and SP acknowledge the support of Indian Space Research Organisation (ISRO) under {\it AstroSat} archival Data utilization program (No. DS\_2B-13013(2)/1/2022-Sec.2). ACP also thanks Inter University centre for Astronomy and Astrophysics (IUCAA), Pune, India for providing facilities to carry out his work.
\bibliographystyle{apj}
\bibliography{3201}

\balance
\end{document}